\documentclass[11pt]{article}
\usepackage{graphicx}
\usepackage{dcolumn}
\usepackage{bm}
\usepackage{amssymb}
\usepackage[pdftex]{color}
\usepackage{float}
\usepackage{lineno,hyperref}
\usepackage{epsfig}
\usepackage{booktabs} 
\usepackage{array} 
\usepackage{paralist}
\usepackage[english]{babel}
\usepackage{xcolor}

\usepackage{geometry} 
\usepackage{amsmath}

\usepackage{subfig} 
\usepackage{fancyhdr} 
\usepackage[titles,subfigure]{tocloft} 

\newcommand {\Real} {\mathbb{R}}	
\newcommand {\bydef}{\,\raise.07485ex\hbox{:}\kern-.025em\hbox{=}\,}
\newcommand{\FT}{\Fb^{\!\top}\!}

\newcommand{\RT}{\Rb^{\!\top}\!}

\newcommand{\QT}{\Qb^{\!\top}\!}

\newcommand{\0} {\textbf{0}}
\newcommand{\Lin} {\mathbb{L}\mathtt{in}}

\newcommand{\Skw}{{\mathbb{S}\textrm{kw}}}

\newcommand{\PSym}{{\mathbb{S}\textrm{ym}}^+}
\newcommand{\Rot} {\mathbb{R}\mathtt{ot}}

\newcommand{\skw} {\mathrm{skw}\,}

\newcommand {\Bc}  {\mathcal{B}}

\newcommand {\Ec}  {\mathcal{E}}

\newcommand {\Hc}  {\mathcal{H}}

\newcommand {\Tc}  {\mathcal{T}}

\newcommand {\Vc}  {\mathcal{V}}
\newcommand {\Wc}  {\mathcal{W}}
\newcommand {\eb} {\mathbf{e}}

\newcommand {\mb} {\mathbf{m}}

\newcommand {\ub} {\mathbf{u}}

\newcommand {\wb} {\mathbf{w}}
\newcommand {\zb} {\mathbf{z}}
\newcommand {\Ab} {\mathbf{A}}
\newcommand {\Bb} {\mathbf{B}}
\newcommand {\Cb} {\mathbf{C}}

\newcommand {\Fb} {\mathbf{F}}

\newcommand {\Hb} {\mathbf{H}}

\newcommand {\Ib} {\mathbf{I}}

\newcommand {\Qb} {\mathbf{Q}}
\newcommand {\Rb} {\mathbf{R}}
\newcommand {\Sb} {\mathbf{S}}
\newcommand {\Tb} {\mathbf{T}}
\newcommand {\Ub} {\mathbf{U}}
\newcommand {\Vb} {\mathbf{V}}

\newcommand {\Wb} {\mathbf{W}}
\newcommand {\Zb} {\mathbf{Z}}
\newcommand {\Do} {\mathbb{D}}
\newcommand {\Eo} {\mathbb{E}}

\newcommand{\Div} {\mathrm{Div}\,}

\newcommand{\asm}{\text{\bfseries\slshape a\/}}

\newcommand{\Asm}{\text{\bfseries\slshape A\/}}
\newcommand{\aref}{\boldsymbol{\mathsf{a}}_0}
\newcommand{\Aref}{\boldsymbol{\mathsf{A}}_0}
\newcommand{\arem}{\asm}

\newcommand{\Arem}{\Asm}

\newcommand{\Sigmab}{\boldsymbol\Sigma}

\newcommand{\tr}{\text{tr}\,}

\begin{document}

\title{Torque--induced reorientation in active fibre-reinforced materials}
\author{Jacopo Ciambella\footnote{jacopo.ciambella@uniroma1.it} 
and Paola Nardinocchi\footnote{paola.nardinocchi@uniroma1.it}\\
{\small \textit{Dipartimento di Ingegneria Strutturale e Geotecnica,
Sapienza Universit\`a di Roma}}}
\date{}

%
%
\maketitle
\begin{abstract}
We introduce a continuum model for a fibre reinforced material in which the reference orientation of the fibre may evolve with time, under the influence of external stimuli. The model is formulated in the framework of large strain hyperelasticity and the kinematics of the continuum is described by both a position vector and by a remodelling tensor which, in the present context, is an orthogonal tensor representing the fibre reorientation process. 
By imposing suitable thermodynamical restrictions on the constitutive equation, we obtain an evolution equation of the remodelling tensor governed by the Eshelby torque, whose stationary solutions are studied in absence of any external source terms. It is shown that the fibres reorient themselves in a configuration that minimises the elastic energy and get aligned along a direction that may or may not be of principal strain. The explicit analysis of the Hessian of the strain energy density allows us to discriminate among the stationary solutions, which ones are stable. 
Examples are given for passive reorientation processes driven by applied strains or external boundary tractions. 
Applications of the proposed theory to biological tissues, nematic or magneto-electro active elastomers are foreseen.
\end{abstract}


%


%
\section{Introduction}
%
In continuum physics, the term \emph{remodelling} is used to denote a large variety of mechanical processes that occur in living tissues \cite{Taber:1996,Cowin:2004,Garikipati:2006,Alford:2007, Criscione:2007,Nagel:2012}, or in active materials \cite{NematNasser:2002,Galante:2013}: as a result of an external stimuli, these materials can actively modify their internal structure to adapt to the new loading conditions. In the framework of continuum mechanics, this effect is modelled by prescribing how an infinitesimal volume element would deform if isolated by the rest of the body \cite{Maugin:1995,Gurtin:2000,Dicarlo:2002} (see also \cite{Tiero:2016}). If neighbouring elements are transformed in a incompatible way, the body maintain its integrity by developing internal stresses \cite{Rodriguez:1994,Nardinocchi:2012,Efrati:2013}; incompatibility may also be viewed as a source of macroscopic instability leading to a sudden change of shape \cite{Pezzulla:2016,Goriely:2017,Aharoni:2017}. 

A particular form of remodelling is reorientation. Active fibre reinforced materials may be able to organise their internal structure and change the orientation of the fibres in response to environmental changes. Many biological tissues have this feature and drive orientation and reorientation by controlled stimuli at chemo-mechanical levels. One paramount example are the osteons in bones, which are able to precisely control the orientation of their internal fibres to optimally tune the mechanical functions during the development; later, during lifetime, fibre reorientation is used to adapt to environmental stimuli \cite{Neville:1993,Driessen:2003,Cardamone:2009}.

Reorientation also occurs in active materials such as liquid crystals, nematic or magneto--rheological elastomers, to cite but a few. Liquid crystals possess some properties typical of liquids as well as certain crystalline properties, that allow them to be described, from a continuum mechanics perspective, as transversely isotropic materials endowed with a characteristic director field representing the macroscopic manifestation of the anisotropy at molecular level. The peculiar property of liquid crystals is that the orientation of the director is easily controlled by magnetic or electric fields due to the anisotropic susceptibility of the molecules \cite{Ericksen:1991,deGennes:1995}.  In Ref.~\cite{Sebastian:2018},   the reorientation of the magnetization by external magnetic fields drives the reorientation of the director field, and the static response as well as the complex dynamics of ferromagnetic liquid crystals under the application of an external magnetic field is investigated. 
Likewise, liquid crystal elastomers present a strong coupling between molecular orientation and macroscopic shape, given by their hybrid character between liquid and rubber. One of their distinctive features is that changes of the molecular orientational order, induced by temperature variations, drive the macroscopic deformation of the body. If the director field changes uniformly within the body, the resulting macroscopic deformation is uniform; inhomogeneous deformations, such as bending or torsion, occur when the director field is nonuniform \cite{DeSimone:2007,Fukunaga:2008,Sawa:2010}. 
Finally, it has been recently shown that short carbon fibres can be aligned during the curing phase of an elastomer by applying an external magnetic field \cite{Stanier:2016}: due to their anisotropic susceptibilities, fibres rotate and dispose parallel to the magnetic field lines. Inhomogeneous field can also be used to achieve complex fibre patterns \cite{Ciambella:2017}. Once the curing process is completed, the movement of the fibres is hampered by the matrix and the application of the magnetic field produces a deformation of the entire solid \cite{Ciambella:2017_PRSA}.

The modelling of reorientation is a major challenge in continuum physics. In the framework  set forth in Ref.~\cite{Dicarlo:2002}, fibre reorientation can be described by an evolution equation derived  by combining suitable constitutive prescriptions with balance equations, and driven by source terms that account for the physical context of the problem. In Ref.~\cite{Dicarlo:2006}, bone tissues were modelled as microstructured continua, whose macroscopic mechanical properties were described by a linear, anisotropic elastic constitutive equation whose moduli evolved with time due to the evolution of the fibre direction. The rotation field that controls the current orientation of the fibres was given as the solution of a first order differential equation whose source term is independent of any external stimuli; thus, this process can be denoted as \emph{passive}. The authors introduced a unified approach to both phenomena (elastic moduli and fibre orientation evolution), based on the principle of virtual powers accompanied by a carefully tailored version of the dissipation inequality, but their model was restricted to small strains and two--dimensional problems.
A different physical context drove the study of deformations induced by the reorientation in nematic elastomers in Ref.~\cite{DeSimone:2007}. The modelling was embedded into the same formal setting of Ref.~\cite{Dicarlo:2006}, but the finite rotation was controlled by an external source and therefore the process was \emph{active}.
Reorientation in hyper--elastic materials was also the focus of the work presented in Ref.~\cite{Himpel:2008}, where a finite element framework was introduced. In that work, the evolution law of the rotation field was postulated independently of the standard mechanical balance, in a way that the fibres reorient in the direction of maximum principal strain. An additional entropy term was added to the dissipation inequality to satisfy thermodynamical consistency. 
A similar approach was used in Ref.~\cite{Melnik:2013} to study fibres alignment and reorientation in biological tissues. The coupling between mechanical anisotropy and tissue remodelling was described by using a transversely isotropic constitutive equation in which the preferred orientation of the fibres evolves towards the direction of maximum principal stretch, following a governing equation postulated by the authors. A strain-driven evolution equation was also used in Refs.~\cite{Driessen:2003,Menzel:2005,Kuhl:2005}. On the contrary, remodelling controlled by the maximum principal stress was used, for instance, in Refs.~\cite{Garikipati:2006,Hariton:2007,Driessen:2008}.

Within the fecund context of growth and remodelling, we aim to  introducing a model, consistent with thermodynamics and finite elasticity, capable of describing the active and passive reorientation in fibre-reinforced materials under a finite deformations regime, within the axiomatic context proposed in  Ref.~\cite{Dicarlo:2002}. To do so, we  use the enriched continuum description represented by a macroscopic deformation field complemented with a time dependent rotation field, tracking the evolution of the fibres. Such a choice allows us to introduce an inner and an outer remodelling torque that are the work conjugates of the reorientation velocity.  Remarkably, it is shown that the inner couple is the skew-symmetric part of the Eshelby stress tensor. The evolution equation of the rotation field is derived by imposing suitable thermodynamical restrictions on the constitutive equations and not postulated a-priori. The solutions of the evolution equation are studied thoroughly with particular emphasis on the passive case, when the reorientation is driven purely by mechanics. It is also shown that the proposed approach can encompass either the linear model in Ref.~\cite{Dicarlo:2006}, and the nonlinear approaches followed in the literature (\cite{Driessen:2003,Menzel:2005,Himpel:2008,Melnik:2013}, to cite but a few), in which the evolution equations were postulated rather than derived from thermodynamics. Moreover, in the case of two equal principal stretches, our model predict that no reorientation takes place, whereas the same condition was forced a-posteriori on the evolution equation in Refs.~\cite{Himpel:2008,Melnik:2013}. Finally, an in-plane remodeling process, in which the fibres lie in a plane orthogonal to the rotation axis, is explicitly studied and serves as a comparison with other literature contributions (see, for instance, Refs.~\cite{Himpel:2008,Melnik:2013,Grillo:2018}). A few examples are given for passive reorientation processes under applied strains or boundary tractions. The evolution of the fibre orientation, as well as the stress and the characteristic times of the remodelling process are analysed through numerical  examples.

\section{Modeling reorientation in finite elasticity}
We set the reorientation modeling within the general framework of the theory of finite elasticity with remodeling presented in Ref.~\cite{Dicarlo:2002} (see also Ref.~\cite{Dicarlo:2006} where fibres reorientation is studied within the context of the linear theory of elasticity). We limit our analysis to transversely isotropic materials and stress the thermodynamic restrictions that make the evolution of fibre orientation coupled with the standard balance equations of finite elasticity.
\subsection{Kinematics}
We define $X$ as the material point of the body, identified with a region $\Bc$ of the Euclidean three-dimensional space $\Ec$. The deformation of the body is a time-dependent map $p :\Bc\times T\to\Ec$ that assigns at each point $X\in\Bc$ a point $x=p(X,t)$ at any instant $t$ of the time interval $T$; accordingly the set $\Bc_t=p(\Bc,t)$ is the configuration of the body at time $t$. We introduce the displacement field $\ub$ such that $x=X +\ub(X,t)$, and we assume that at some instant $t=t_0$, the body occupies the reference configuration, that is,  $p(X,t_0)=X$ for each point in $\Bc$. We consider deformations that are twice continuously differentiable and, so, we can write 
\begin{equation}
\Fb(X,t)=\nabla\,p(X,t)\qquad \text{and} \qquad \dot{p}(X,t)=\dfrac{\partial p}{\partial t}(X,t)\,,
\end{equation}
that represent the deformation gradient and the material velocity field, respectively. In addition, 
we introduce a material vector field $\aref:\Bc \to\Vc$, with $\Vc$ the  translation space of $\Ec$, such that $\aref\cdot \aref=1$, that represents the preferred direction that the internal structure endows to the material, i.e., the direction of the fibre at position $X$. The corresponding orientation tensor is $\Aref = \aref\otimes\aref$. 

The remodeling process is defined by a time-dependent rotation, here identified with a tensor field ${\Rb:\Bc\times\!T \to \Rot}$, that makes the referential orientation $\Aref(X)$ evolving with time; accordingly, we write
\begin{equation}
\Arem(X,t)=\Rb(X,t)\,\Aref(X)\Rb(X,t)^T\,,
\end{equation}
for the remodeled orientation tensor, with $\Arem=\arem\otimes\arem$, and
\begin{equation}
\arem(X,t)=\Rb(X,t)\,\aref(X)\,,
\end{equation}
for the remodelled fibre orientation.
In the present framework, we call the pair $(p,\Rb)$ the local configuration space and the associated velocity field is identified by the time derivatives $(\dot{p},\dot\Rb\RT)\in\Vc\times\Skw$. As such, the pair $(\wb,\Wb)\in\Vc\times\Skw$ are the corresponding virtual velocities. 

Throughout this paper, the dependence on the referential coordinates $X$ and the time $t$ will left tacit unless otherwise specified.

\subsection{Balance equations}

In this section, through the principle of null working, we introduce the balance equations of forces and the further balance equation which drives the evolution of the remodelling tensor $\Rb$.

By introducing the forces and torques that act conjugate to each kinematic variable, we define the working, which is a continuous, linear, real-valued functional on the space of virtual velocities, that can be additively decomposed into an outer virtual working $\Wc_e$ and an inner virtual working $\Wc_i$. Therefore, we write
\begin{equation}
\Wc_e(\wb,\Wb)=\int_\Bc(\zb\cdot\wb + \Zb\cdot\Wb) + \int_{\partial\Bc}\mathbf s\cdot\wb\quad\textrm{and}\quad
\Wc_i(\wb,\Wb)=\int_\Bc(\Sb\cdot\nabla\wb + \Sigmab\cdot\Wb)\,.
\end{equation}
The pair $(\zb,\mathbf s)$ are the forces per unit of (referential) volume and  area, respectively, and, consequently, $\Sb$ is the (first) Piola--Kirchhoff stress tensor. The pair of tensors $(\Zb,\Sigmab)$ are the torques per unit of (referential) volume and area, and represent the outer and inner remodelling torques. The outer torque $\Zb$ is an external source term, that can incorporate stimuli such as the mechanical effects of the biochemical control system in biological tissues or the thermal and electric field-induced effects observed in nematic elastomers \cite{Taber:1996,Fukunaga:2008}. 

By enforcing the condition that the working must be null for any test velocities $(\wb,\Wb)\in\Vc\times\Skw$, we obtain the following balance equations and boundary conditions:
\begin{equation}\label{be}
\Div\Sb + \zb =\0\quad\textrm{and}\quad\Sigmab=\Zb\,\,\,\textrm{in}\,\,\Bc\,,
\end{equation}
\begin{equation}\label{bcmec}
\ub=\hat\ub\,\,\,\textrm{in}\,\,\partial_u\Bc \quad\textrm{and}\quad\Sb\,\mb=\mathbf{s}\,\,\,\textrm{on}\,\,\partial_t\Bc\,,
\end{equation}
with $\partial_u\Bc$ and $\partial_t\Bc$ as the parts of the boundary $\partial\Bc$ where displacements and tractions are prescribed and $\mb$ the unit normal to $\partial_t\Bc$. The actual outer working, that corresponds to balanced forces and torques $(\zb,\mathbf{s};\Zb)$, can be reduced to the form 
\begin{equation}
\Wc_e(\dot p,\dot\Rb\RT) =\int_\Bc(\zb\cdot\dot p + \Zb\cdot\dot\Rb\RT) + \int_{\partial_t\Bc}\mathbf{s}\cdot \dot p=\int_\Bc(\Sb\cdot\dot\Fb + \Sigmab\cdot\dot\Rb\RT)\,,
\end{equation}
which identifies the power expended during the evolution of the continuum. 

\subsection{Constitutive Equations}

We now consider a class of materials which admits a strain-energy density function which is, in general, a function of $\Fb$. However, the existence of a preferred direction makes the material response transversely isotropic and implies that the strain energy function additionally depends on the evolution of the anisotropy axis. This, together with the requirement of invariance under both changes of observer and rigid body modifications, allows the strain energy density to be written as the map $\psi: \PSym\!\times \Rot  \mapsto \Real$ given by
\begin{equation}
\psi( \Cb,\Rb)= \phi(\Cb, \Rb\, \Aref\,\RT)=\phi(\Cb,\Arem)\,,
\label{StrainEnergy}
\end{equation}
in which the function $\phi$ is an isotropic function of the right Cauchy-Green strain tensor $\Cb=\FT\Fb$ and of the remodelled orientation tensor $\Arem$.\footnote{In writing down  Eq.~\eqref{StrainEnergy}, we have assumed that  the material properties are independent of the sense of $\aref$. A material for which this requirement does not hold is usually referred to as \emph{transversely hemitropic}.}
This entails the symmetry group 
\begin{equation}
\mathcal{G}_t = \lbrace \Qb \in \Rot \vert \phi(\Qb\,\Cb\,\QT , \Qb\,\Arem\, \QT)=\phi(\Cb,\Arem)\rbrace
\end{equation}
of the stored energy function to evolve with time through the remodelled tensor $\Arem$.

It is well-known that for a transversely isotropic material, such as the one described by Eq.~\eqref{StrainEnergy}, the dependence of the strain energy function on the pair $(\Cb,\Arem)$ is through five principal invariants, 
\begin{equation}
I_1=\tr{\Cb}\,,\quad I_2=\frac 12 \left[ (\tr \Cb)^2-\tr\Cb^2\right]\,,\quad I_3=\det\Cb\,,
\label{I1I3}
\end{equation}
and\begin{equation}
I_4 = \Cb\cdot\Arem\,,\quad I_5=\Cb^2\cdot\Arem\,,
\label{I4I5}
\end{equation}
which depend on both the strain and, through $\Arem$, the remodelling tensor $\Rb$ (as it is the case for $I_4$ and $I_5$). In this sense, the invariant  $I_4$ has the direct interpretation as the square of the stretch in the remodelled direction of the fibre, whereas the invariant $I_5$ introduces an additional effect that relates to the behavior of the reinforcement under shear deformations \cite{Merodio:2005}. Equations \eqref{I1I3}-\eqref{I4I5} allow us to write the strain energy function in \eqref{StrainEnergy} as
\begin{equation}
\phi(\Cb,\Arem)=\hat\phi(I_1,I_2,I_3,I_4,I_5)\,.
\label{energyinvariants}
\end{equation}
\subsection{Thermodynamics restrictions}\label{SectTherm}
In the mechanical context we consider, we replace the first and second laws of thermodynamics by an energy imbalance \cite{Coleman:1974}, which expresses the requirement that, for any realizable process $(p,\Rb)$, the time-rate of the total energy does not exceed the outer working expended along the same process. So, the (local form of the) dissipation inequality yields
\begin{equation}\label{di}
\dot\psi \le \Sb\cdot\dot\Fb + \Sigmab\cdot\dot\Rb\RT\,,
\end{equation}
along any process and for each time instant. In view of \eqref{StrainEnergy}, the time rate of strain energy density is rewritten in terms of partial derivatives as
\begin{equation}
\dot{\psi}=\frac{\partial \psi}{\partial \Cb}\cdot \dot{\Cb} + \frac{\partial \psi}{\partial \Rb}\cdot \dot\Rb\,.
\end{equation}
By using trivial properties of the algebra of symmetric tensors, we write
\begin{align}
\dfrac{\partial \psi}{\partial \Cb}\cdot\dot\Cb &= 2\,\Fb\dfrac{\partial \phi}{\partial \Cb}\cdot\dot\Fb\,,\label{comm}\\
\dfrac{\partial \psi}{\partial \Rb}\cdot\dot{\Rb} &=\big[\frac{\partial \phi}{\partial \Arem},\Arem\big]\cdot\dot\Rb\RT \,,\label{comm1}
\end{align}
where we have introduced the commutator operator ${[\cdot,\cdot]:\Lin\times\Lin \to \Skw}$ defined by
\begin{equation}
[\Ab,\Bb]=\Ab\Bb-\Bb\Ab,\quad \forall \Ab,\Bb\in \Lin\,.
\end{equation}
%
With these results in hand, the dissipation inequality \eqref{di} takes the form
\begin{equation}\label{dinew}
\big(-\Sb + 2\Fb\frac{\partial\phi}{\partial\Cb}\big)\cdot\dot\Fb + \big(-\Sigmab + \big[\frac{\partial \phi}{\partial \Arem},\Arem\big]\big)\cdot\dot\Rb\RT \le 0\,.
\end{equation}
Notice that the second term in the inequality is sometimes added artificially, and called extra entropy term, to satisfy thermodynamical consistency and capture the effects related to processes that may stiffen the material~\cite{Himpel:2008}. Here, on the contrary, such a term naturally arises from the  kinematic description.

If there is no elastic dissipation, the stress is completely determined by $\Sb = 2\Fb\partial\phi/\partial\Cb = \partial\phi/\partial\Fb$, whereas the inner remodelling torque can be split into an elastic part $\Sigmab_e$ and a dissipative part $\Sigmab_d$:
\begin{equation}
\Sigmab=\Sigmab_e + \Sigmab_d\quad\textrm{with}\quad\Sigmab_e =[\frac{\partial \phi}{\partial \Arem},\Arem]\quad\textrm{and}\quad\Sigmab_d=\Do(\Cb,\dot\Rb\RT)\,\dot\Rb\RT\,,
\label{torque}
\end{equation}
where $\Do$ is a fourth order positive definite tensor, such that $\Do\Wb\cdot\Wb\ge 0$ for any $\Wb\in\Skw$ and represents the resistance to remodelling. 

To make explicit the dependence of the elastic torque on the stress tensor, we now further exploit the relationship~\eqref{energyinvariants}. We first obtain the expression of the second Piola-Kirchhoff stress tensor $\Tb=\Fb^{-1}\Sb=2\partial\phi/\partial\Cb $ of a transversely isotropic material as
\begin{align}
\Tb &=  2(\phi_1 + I_1 \phi_2)  \Ib - 2 \phi_2 \Cb + 2 I_3 \phi_3 \Cb^{-1}
 +2 \phi_4 \Arem + 2 \phi_5\big( \Arem \Cb + \Cb \Arem\big)\,,
\label{SecondPK}
\end{align}
with $\phi_i=\partial \hat\phi/\partial I_i$ (i=1...5); then, by making use of the derivatives of the strain invariants $\partial I_4/\partial \Arem = \Cb$ and $\partial I_5/\partial\Arem=\Cb^2$, we calculate
\begin{equation}
\dfrac{\partial \phi}{\partial \Arem} = \phi_4 \Cb + \phi_5 \Cb^2\,,
\end{equation}
that allows us to write the following alternative expression of the commutator
\begin{align}
\big[ \dfrac{\partial \phi}{\partial \Arem},\Arem\big] &=[\phi_4\Cb+\phi_5\Cb^2,\Arem]\notag \\
&=[\Cb,\phi_4\Arem+\phi_5(\Arem\Cb+\Cb\Arem)] = \frac 12 [\Cb,\Tb]\,.
\label{comm2}
\end{align}
In \eqref{comm2}, we have used the equivalence  $[\Cb^2,\Arem]=[\Cb,\Arem\Cb+\Cb\Arem]$ and the fact that $[\Cb,\Ib]=[\Cb,\Cb^{-1}]=[\Cb,\Cb]=\0$.  The elastic torque in \eqref{torque} is hence written as
\begin{equation}
\Sigmab_e = \frac 12 \big[\Cb,\Tb\big]\,,
\label{TorqueStress}
\end{equation}
which is the generalisation to the nonlinear settings of the linear remodelling torque introduced in \cite{Dicarlo:2006} and points out how the stress tensor and the remodelling self-torque are coupled. As highlighted in Ref.~\cite{Dicarlo:2002}, this coupling is mandatory and independent of any special assumption on the strain energy. Additional couplings--through the external field $\Zb$, in particular--are legitimate, and may assume different forms depending on the type of stimuli that induce the fibre reorientation.

\paragraph*{{\bf Remark}} The relationship \eqref{TorqueStress} can be further rearranged to make the \emph{Eshelby stress} tensor $\Eo=\phi(\Cb,\Arem) \Ib - \Cb\Tb $ appearing. In fact
\begin{equation}
\skw\Eo=\frac 12 [\Cb,\Tb]=\Sigmab_e\,,
\end{equation}
that justifies the appellative of \emph{Eshelbian} torque used to refer to $\Sigmab_e$.

\subsection{The  remodelling equation}

Using the above derived constitutive equations into the balance Eqs.~(\ref{be}), we obtain the evolution equation which drives the remodelling of the microstructure:
\begin{equation}\label{una}
\Do\dot\Rb\Rb^T = \Zb + \frac 12 [\Tb,\Cb]\quad\textrm{in}\quad\Bc\times\Tc\qquad\text{and}\qquad\Rb=\Ib\quad\textrm{in}\quad\Bc\times\{t_0\}\,,
\end{equation}
where we have assumed that at time $t_0$ the configuration of the body coincides with the reference configuration and that $\Do$ is constant; however, more complex choices can be easily accounted for. Correspondingly, at each instant, we have a family of elastic problems driven by the Eq.~(\ref{una}) and governed by the balance of forces
\begin{equation}\label{due}
\Div\Sb + \zb =\0\quad\textrm{with}\quad
\Sb = \Fb\Tb\,,\quad\textrm{and}\quad \Tb=2\frac{\partial\phi}{\partial\Cb}\,.
\end{equation}
 Reorientation is denoted as passive if $\Zb=\0$; otherwise, the torque $\Zb$ makes up the effect of an external stimulus on the reorientation of the fibres, and such a reorientation is denoted as active.
As an example, we cite Ref.~\cite{DeSimone:2007} where it is assumed that the external source $\Zb$ models the effect of an electric field on the director rotation and can be represented as a function of the field and $\Arem$ depending on the elastomer permittivity and the  Frank constant (see Eq.~(3) in Ref.~\cite{DeSimone:2007}). 

Actually, since the time scales associated with remodeling are much longer than those associated with inertia, no inertial loads will be considered in the balance of forces. Hence, in absence of other bulk loads, we set $\zb=\0$ from now on.

\section{Passive remodelling}
We start by studying the stationary solutions of the remodelling Eq.~\eqref{una} in the purely passive case, i.e., $\Zb=\0$. Equation \eqref{una}$_1$ is rewritten as
\begin{equation}
\Do\dot\Rb\RT = \frac 12 [\boldsymbol{\mathcal{T}}(\Cb,\Rb),\Cb]\,,
\label{PassiveRemodelling}
\end{equation}
where the function $\boldsymbol{\mathcal{T}}(\Cb,\Rb)=\Tb$ is used to indicate the constitutive Eq.~\eqref{SecondPK} and to highlight the dependence of the stress tensor $\Tb$ on the remodelling tensor $\Rb$.

Remarkably, we are going to show that the source term $[\boldsymbol{\mathcal{T}}(\Cb,\Rb),\Cb]$ in \eqref{PassiveRemodelling} is indeed the derivative of the strain energy density $\psi$ with respect to the rotation tensor $\Rb$. This implies that, for any given deformation $\Cb$, the stationary solutions of the remodelling equation, for which $\dot\Rb\RT=\0$, are extremal points of the elastic strain energy, i.e., equilibrium solutions of the elastic problem. 


To prove this property, we necessitate the following definition of \emph{coaxiality}:
given two symmetric tensors $\Ub$ and $\Vb$, we have
\begin{equation}
\Ub,\Vb\;\text{coaxial}\qquad \Leftrightarrow \qquad \Ub\Vb=\Vb\Ub\qquad \Leftrightarrow \qquad [\Ub,\Vb]=\0\,.
\label{DefinitionCoaxiality}
\end{equation} 
Equation~\eqref{DefinitionCoaxiality} highlights that the commutator is indeed a measure of the defect of coaxiality between symmetric tensors, in the same way as the cross product measures the defect of collinearity between vectors.

We are now in the position of introducing the following Proposition that completely characterises the class of stationary solutions of \eqref{PassiveRemodelling}:
\paragraph*{{\bf Proposition 1}}
\textit{
For any given deformation $\Cb$, the following statements are equivalent
\begin{enumerate}[(a)]
\item $\Rb_\ast$ is a stationary solution of the remodelling Eq.~\eqref{PassiveRemodelling};
\item stress $\Tb_\ast=\boldsymbol{\mathcal{T}}(\Cb,\Rb_\ast)$ and strain $\Cb$ tensors are \emph{coaxial};
\item $\Rb_\ast$ is a stationary point of the map 
\begin{equation}
\sigma:\Rot\ni\Rb\mapsto\sigma(\Rb)=\psi(\Cb,\Rb)\,,
\label{sigma}
\end{equation}
where $\psi$ is the strain energy density defined in \eqref{StrainEnergy}.
\end{enumerate}
}
Before passing to the proof, we highlight two important facts: (i) the condition (b) of coaxiality between the symmetric Piola-Kirchhoff stress and the right Cauchy-Green strain is equivalent to the coaxiality between the Cauchy stress $\bm \sigma = (\det\Fb)^{-1}\Fb\Tb\FT$ and the left Cauchy-Green strain $\Bb=\FT\Fb$ (see for instance Ref.~\cite{Vianello:1996a}); (ii) the function $\sigma$ defined in \eqref{sigma} is a continuous function over the compact Lie group $\Rot$ and, hence, has an absolute minimum and maximum; therefore, for each prescribed deformation $\Cb$, there are always at least two stationary solutions.

\paragraph*{Proof}

The equivalence between $(a)$ and $(b)$ is an immediate consequence of the definition of coaxiality \eqref{DefinitionCoaxiality}, for which if $\Tb_\ast$ and $\Cb$ are coaxial, then $[\Tb_\ast,\Cb]=\0$, and, hence, $\Rb_\ast$ is a stationary solution of \eqref{PassiveRemodelling}.

To prove the equivalence between $(b)$ and $(c)$, we follow Ref.~\cite{Vianello:1996a}, and note that the map $\sigma$ is a continuous and derivable function whose stationary points can be characterised by letting the tangential derivative at $\Rb_\ast$ to be zero. By exploiting the canonical isomorphism between the tangent space $\text{Rot}(\Rb)$ at $\Rb\in\Rot$ and $\Skw$ for which $\text{Rot}(\Rb)=\lbrace \Wb\Rb\vert \Wb\in\Skw\rbrace$, we write
\begin{equation}
\dot\sigma(\Rb) = D\sigma(\Rb)[\Wb\Rb] =  -\frac{1}{2} [\boldsymbol{\Tc}(\Cb,\Rb) ,\Cb]\cdot \Wb\,,
\label{sigmaDot}
\end{equation}
in view of the result in Eqs.~\eqref{comm1} and \eqref{comm2}. Equation \eqref{sigmaDot} is zero for every $\Wb$ if and only if $\Tb$ and $\Cb$ are coaxial, thus the equivalence between $(b)$ and $(c)$.\\[0.1cm]

Proposition 1 highlights a major difference between our approach and those followed in Refs.~\cite{Driessen:2003,Menzel:2005,Himpel:2008,Melnik:2013}: in a passive remodelling process, the evolution of the material microstructure is totally controlled by the elastic energy in the sense that the fibre rotates towards a configuration which makes the energy stationary; differently, in other literature contributions the rotation of the fibres is controlled by the maximum principal strain or stress. However, the two approaches are partially reconciled by the following proposition.

\paragraph*{{\bf Proposition 2}}
\textit{The rotation $\Rb_\ast$ that aligns the remodelled preferred orientation ${\arem_\ast=\Rb_\ast\,\aref}$ with a principal strain direction, is a stationary solution of \eqref{PassiveRemodelling}.}  

\paragraph*{Proof} If the remodelled orientation $\arem_\ast$ coincides with a principal strain direction, by using the constitutive Eq.~\eqref{SecondPK} and by calling $\lambda_\ast$ the associated principal stretch (eigenvalue of $\Cb$), we write
\begin{equation}
\Tb_\ast\,\arem_\ast = 2(\phi_1+I_1\phi_2) \arem_\ast -2\phi_2\lambda_\ast\arem_\ast + 2I_3 \phi_3\lambda_\ast^{-1}\arem_\ast + 2 \phi^\ast_4 \arem_\ast + 4 \lambda_\ast \phi^\ast_5 \arem_\ast
\label{RemodStress}
\end{equation}
where the symbols $\phi^\ast_i=\partial\phi_i(\Cb,\Rb_\ast \Aref \Rb_\ast^{\!\top})/\partial I_i$, ($i=4,5$), are used to highlight the dependence of the energy derivatives on the rotation $\Rb_\ast$. Equation~\eqref{RemodStress} shows that $\arem_\ast$ is also a principal direction of stress, thus, $\Tb_\ast$ and $\Cb$ are coaxial and in view of the result in Prop. 1, $\Rb_\ast$ is a stationary solution of the evolution Eq.~\eqref{PassiveRemodelling}.

\paragraph*{{\bf Remark}} Due to the coaxiality between $\Tb_\ast$ and $\Cb$ at the stationary solution, the principal directions of stress and strain coincide, thus, Proposition 2 could have been equivalently formulated in terms of \emph{principal stresses}.\vspace{0.2cm}

Proposition~2 highlights the fact that the rotations which make the fibres aligned with the principal strain directions are stationary solutions of the remodelling equation. However, the converse is in general not true and there may be other rotations which solve \eqref{PassiveRemodelling}. In fact, if we call $\arem_\ast$ a remodelled fibre orientation which is not a principal direction of strain, and we indicate with $\eb$ the actual principal strain direction and with $\lambda$ the associated principal stretch, we have
\begin{equation}
\Tb_\ast \eb=  2(\phi_1 + I_1 \phi_2)  \eb - 2 \phi_2 \lambda \eb + 2 I_3 \phi_3 \lambda^{-1}\eb
 +2 \phi_4^\ast (\arem_\ast\cdot \eb)\,\arem_\ast + 2 \phi_5^\ast (\arem_\ast\cdot \eb)\big( \lambda\, \arem_\ast+ \Cb \arem_\ast\big)\,,
 \label{StressStar}
\end{equation}
with $\arem_\ast\cdot \eb\neq 0$. Equation \eqref{StressStar} shows that $\Tb_\ast$ and $\Cb$ are coaxial, and $\Rb_\ast$ is a stationary solution, if and only if $\phi_4^\ast$ and $\phi_5^\ast$ vanish, i.e.,
\begin{equation}
\phi_4^\ast(\Cb,\Rb_\ast \Aref \Rb_\ast^{\!\top})=\phi_5^\ast(\Cb,\Rb_\ast \Aref \Rb_\ast^{\!\top})=0.
\label{implicitR}
\end{equation}
These two equations define two surfaces on the space of rotation $\Rot$, whose intersection gives the rotations $\Rb_\ast$ that solve Eq.~\eqref{PassiveRemodelling}. If these rotations exist, they select  remodelled orientations $\arem_\ast$ along which fibres are totally unstretched. This together with the assumption that the strain energy is convex in $I_4$ and $I_5$ guarantees that the corresponding configuration is a minimum of the elastic energy. 

The existence of solutions of \eqref{implicitR} clearly depends on the  deformation $\Cb$, as it will be shown in the next section for some examples. In any case, we ought to remark that this additional solution is not caught by the maximum principal strain approach followed in the literature.

To further assess the stability of the stationary solutions and give a precise criterion for deciding whether they are a  minimum or a maximum of the energy, we now evaluate the second derivative of $\sigma$, i.e., the Hessian of the strain energy density. If the Hessian in correspondence of the stationary solution is positive definite, then the energy is convex and, due to the minus sign in Eq.~\eqref{sigmaDot}, the associated rotation is a stable solution of the evolution equation. This again confirms that Eq.~\eqref{PassiveRemodelling} drives the remodelling process towards a fibre orientation that minimise the elastic energy.

%
To evaluate the second derivative of $\sigma$, we make use of \eqref{sigmaDot} and, by fixing $\Ub\in \Skw$, we consider the smooth function $\Delta:\Rb\mapsto D\sigma(\Rb)[\Ub\Rb]$ defined on $\Rot$. The tangential derivative of $\Delta$ at $\Rb$ in the direction corresponding to the skew-symmetric tensor $\Vb\in\Skw$ is
\begin{equation}
\dot\Delta= D\Delta(\Rb)[\Vb\Rb]= \big [\dfrac{\partial \phi}{\partial \Arem},\Ub\big] \cdot[\Vb,\Arem] + 2 \dfrac{\partial^2 \phi}{\partial^2 \Arem}[ \Ub\Arem]\Arem\cdot \Vb\,,
\label{hes0}
\end{equation}
that, after using Eq.~\eqref{energyinvariants}, and exploiting the algebraic properties of skew-symmetric tensors, can be written as the symmetric bilinear form $\Hc[\Ub,\Vb]:\Skw\times\Skw \mapsto R$  defined by
\begin{align}
\dot\Delta=\Hc[\Ub,\Vb] = &[\phi_4\Cb + \phi_5\Cb^2,\Ub]\cdot [\Vb,\Arem] 
+2\,\phi_{44}(\Cb\Arem\cdot \Ub)\,(\Cb\Arem\cdot \Vb)\notag \\
&+2\,\phi_{55}(\Cb^2\Arem\cdot \Ub)\,(\Cb^2\Arem\cdot \Vb)\notag\\
 &+2\,\phi_{45}\big[(\Cb\Arem\cdot \Ub)\,(\Cb^2\Arem \cdot \Vb)+(\Cb\Arem\cdot \Vb)\,(\Cb^2\Arem \cdot \Ub)\big]\,,
 \label{hes}
\end{align}
where $\phi_{ij}=\partial^2 \hat \phi/\partial I_i\partial I_j$.\\

The evaluation of the sign definiteness of the Hessian can be carried out by exploiting the well-known isomorphism between the skw-symmetric tensors and the Euclidean three-dimensional vector space. Upon choosing a proper basis in $\Skw$, one can identify $\Hc$ with a second order symmetric tensor $\Hb$ whose components are $H_{ij} = \Hc[\Wb_i,\Wb_j]$; accordingly, the evaluation of the sign definiteness of $\Hc$ is reduced to the study of the sign of $\det(\Hb)$. A detailed analysis of the energy convexity in correspondece of the stationary solutions will be carried out in the next section.


\section{In-plane remodeling}
\label{ipr}
In this Section we study the case of fibres that, at each time instant, lie in a plane orthogonal to the axis of rotation and we refer to it as \emph{in-plane remodelling}. We call $\lbrace \eb_1,\eb_2,\eb_3 \rbrace$ an orthonormal basis in the vector space $\Vc$ and assume that the plane oriented with $\eb_3$ contains the fibres and the rotation has an axis parallel to $\eb_3$ ($\Rb\eb_3=\eb_3$). 
To  further simplify the analysis, we assume that the mobility tensor $\Do$ is isotropic and determined by the positive scalar constant $m$ as $\Do=m/2\,\Ib$. The rotation tensor $\Rb$ is parametrized with the angle $\theta=\theta(X,t)$, that represents the remodelled direction that the fibre forms with the $\eb_1$-axis, i.e, $\arem=\cos\theta\,\eb_1+\sin\theta\,\eb_2$.  Accordingly, the initial orientation field $\Aref$ is determined by the angle $\theta(X,0)=\theta_0(X)$, whereas the stationary orientation field $\Arem_\ast$  by the angle $\theta(X,\infty)=\theta_\ast(X)$. With this parametrisation, the evolution Eq.~\eqref{una} reduces to a scalar equation of the sole variable $\theta$. In fact, the rate of change of the remodelled direction $\dot \arem = \dot\Rb\RT\,\arem$ can be rewritten as
\begin{equation}\label{unap}
\dot\arem = \frac{1}{m}[\Tc(\Cb,\Rb),\Cb]\arem\,,
\end{equation}
or, by using  the constitutive Eq.~\eqref{SecondPK}, as
\begin{equation}\label{unap2}
\dot\arem = \frac{1}{m}\left(\phi_4(I_4\Ib-\Cb)+\phi_5(I_5\Ib-\Cb^2)\right)\arem\,.
\end{equation}
From here, by assuming that $\eb_1$ and $\eb_2$ are the principal strain directions\footnote{If this were not the case, then a rotated reference system could always be used to have the axes coincide with the principal strain directions, yet the remodelling Eq.~\eqref{PassiveRemodelling} would give the rotation with respect to these new reference system.} and by using the parametrisation of $\arem$ in terms of $\theta$, we arrive at
\begin{align}
\dot\theta = \frac{1}{2m}(\lambda_1-\lambda_2)\big(\phi_4(\lambda_1,\lambda_2,\theta)+(\lambda_1+\lambda_2)\,\phi_5(\lambda_1,\lambda_2,\theta)\big)\sin(2\theta),
\label{1dremod}
\end{align}
where we have kept explicit the dependence of the energy derivatives $\phi_4$ and $\phi_5$ on the principal stretches $\lambda_1$, $\lambda_2$ and the remodelled angle $\theta$. 

Firstly, Eq.~(\ref{1dremod}) immediately shows that  if $\lambda_1=\lambda_2$, $\dot\theta=0$ and no remodelling occurs. Actually, this result can be also inferred from Eq.~\eqref{unap}; in fact,  if the two principal stretches associated to the directions $(\eb_1,\eb_2)$ are the same, the commutator $[\Tb,\Cb]$ vanishes for every initial orientation of the fibre. Indeed,  if 
\begin{align}
\Cb&=\lambda_1 \,\eb_1\otimes\eb_1 +\lambda_2\,\eb_2\otimes\eb_2 + \lambda_3 \,\eb_3\otimes\eb_3\,,
\end{align}
and $\lambda_1=\lambda_2=\lambda$, then, from Eq.~\eqref{SecondPK}, we get 
\begin{equation}
\big[\Tb,\Cb\big] = 2 \lambda(\phi_4+\lambda \phi_5)\big( \arem\cdot\eb_1 (\arem\otimes \eb_1 -\eb_1\otimes \arem) + \arem\cdot\eb_2 (\arem\otimes \eb_2 -\eb_2\otimes \arem)\big)=\0\,,
\label{comzero}
\end{equation}
due to the properties of the dyadic product.\footnote{Notice that for the in-plane remodeling process considered, all the results are independent of $\lambda_3$.} As such, $\dot\Rb\RT=\0$, $\Rb(X,t)=\Ib$ and $\arem(X,t) = \aref(X)$ for all $t$, i.e., the fibre maintains its reference orientation for every value of the principal stretches.  It is worth noting that in the maximum principal strain approach followed in the literature, this null remodelling condition, which naturally arises in the present theory, is usually imposed a-posteriori on the evolution equation.

For $\lambda_1\neq \lambda_2$ and $-90^\circ < \theta \leq 90^\circ$, Eq.~ \eqref{1dremod} can have several stationary solutions: $\theta_\ast=0^\circ$ and $\theta_\ast =90^\circ$ are the solutions associated to the principal strain directions and exist for every value of the principal stretches; additional solutions may exist if the condition \eqref{implicitR} is satisfied.
In particular, when the anisotropic part of the strain energy function is proportional to $
(I_4-1)^2$ and independent of $I_5$, we have\footnote{
As usually assumed in the literature (see for instance \cite{Melnik:2013}).}
\begin{equation}
\phi_4 = \frac{1}{2}\mu \,\gamma \left(\lambda_1+\lambda_2+(\lambda_1-\lambda_2)\cos(2\, \theta)-2\right),\qquad \phi_5=0\,,
\label{phi4phi5}
\end{equation}
with $\mu$ and $\gamma$ constitutive parameters ($\mu,\gamma>0$). Accordingly,  Eq.~\eqref{implicitR} has only one solution when when $\lambda_1$ is tensile and $\lambda_2$ is compressive or viceversa. 

\begin{figure}
\begin{center}
\begin{tiny}
 \def\svgwidth{\textwidth}
 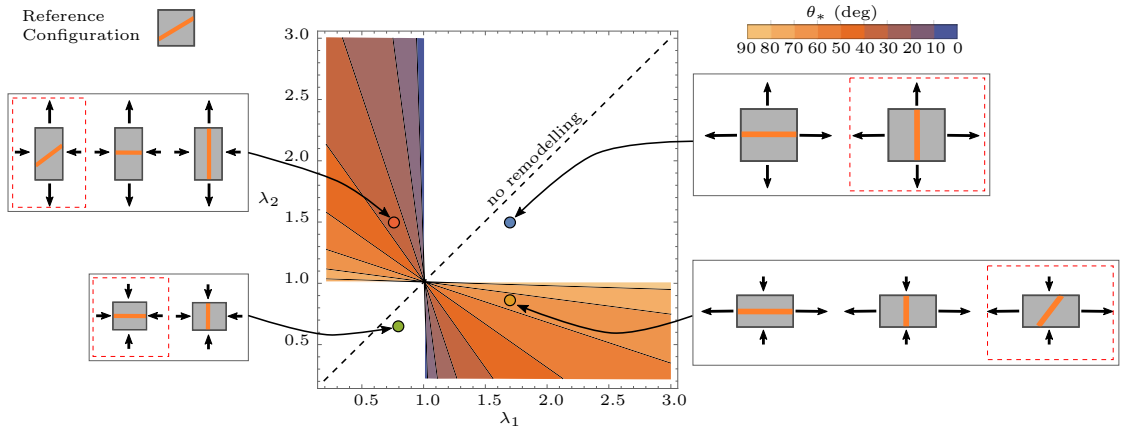
\end{tiny}
\end{center}
\caption{Stationary solutions of the passive remodelling Eq.~\eqref{1dremod} for different values of $\lambda_1$ and $\lambda_2$: in the white region, only two solutions exist, namely $\theta_\ast=0^\circ$ and $\theta_\ast=90^\circ$, whereas the contour plot shows the additional solution in terms of the remodelled fibre angle $\theta_\ast$. The insets show the configurations of a representative volume element for the stretch values marked with dots in the figure; stable solutions are enclosed by a dashed line. For $\lambda_1=\lambda_2$, no-remodelling occurs.}
\label{fig:cp}
\end{figure}

The entire set of the stationary solutions of \eqref{1dremod} is plotted in Fig.~\ref{fig:cp} in terms of the stationary angles $\theta_\ast$. 
It is seen that, when both the principal stretches are tensile or compressive, Eq.~\eqref{implicitR} has no solution and, therefore, only the stationary solutions corresponding to the principal strain directions are found. On the contrary, if one of the stretches is tensile and the other compressive, the additional solution corresponding to the remodelled fibre angle plotted in the figure is found. Finally, if $\lambda_1=\lambda_2$, no remodelling occurs and the fibres maintain the initial orientation (dashed line in the figure).

\begin{figure}
\begin{center}
\begin{tiny}
 \def\svgwidth{\textwidth}
 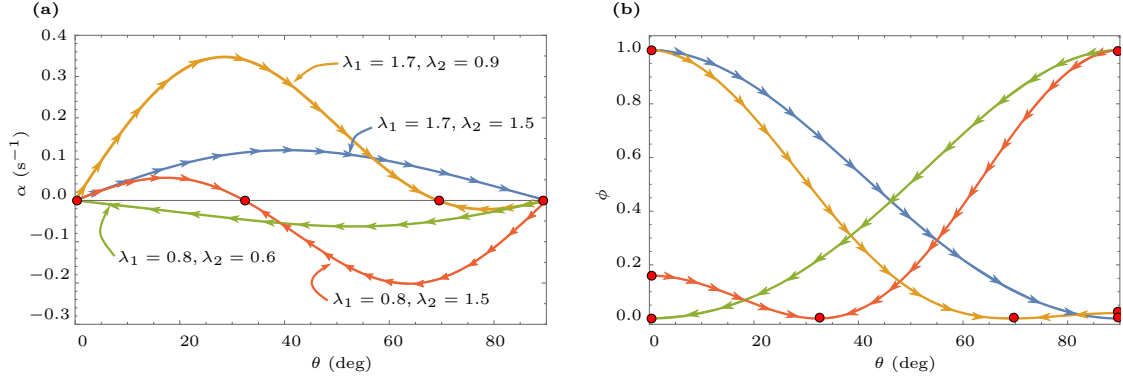
\end{tiny}
\end{center}
\caption{(a) Plot of the function $\alpha(\theta)$ for the different values of the principal stretches $\lambda_1$ and $\lambda_2$.  Stationary solutions of the remodelling Eq.~\eqref{1dremod} correspond to the angles at which the curves intersect the horizontal axis. (b) Corresponding anisotropic part of the elastic energy normalised between $0$ and $1$.}
\label{fig:alpha}
\end{figure}

The stability of the stationary solutions is studied by analysing the source term, i.e., the right hand side of the Eq.~\eqref{1dremod}, which, for given $\lambda_1$ and $\lambda_2$, is a function of $\theta$ defined by
\begin{equation}\label{alpha}
\alpha(\theta) = \frac{1}{2m}(\lambda_1-\lambda_2)\phi_4(\lambda_1,\lambda_2,\theta)\sin(2\theta)\,,
\end{equation}
whose graph, with $\phi_4$ given by \eqref{phi4phi5}, is shown in Fig.~\ref{fig:alpha} for different values of $\lambda_1$ and $\lambda_2$. As expected from previous analysis, when $\lambda_1$ and $\lambda_2$ are both tensile (compressive) only two solutions are found. In particular, when $\lambda_1>\lambda_2>1$ (blue curve), $\alpha'(0^\circ)>0$ and the solution $\theta_\ast=0^\circ$ is unstable, whereas $\alpha'(90^\circ)<0$ and $\theta_\ast=90^\circ$ is a stable solution. On the contrary, when $\lambda_2<\lambda_1<1$ (green curve), $\alpha'(0^\circ)<0$ and the solution $\theta_\ast=0^\circ$ is stable, whereas $\alpha'(90^\circ)>0$ and $\theta_\ast=90^\circ$ is unstable. When the third solution appears (orange and yellow curves), the same reasoning shows that $\theta_\ast=0^\circ$ and $\theta_\ast=90^\circ$ are unstable solutions, whereas the one corresponding to the condition \eqref{implicitR} is stable, for every $\lambda_1$ and $\lambda_2$. In this configuration, fibres are unstretched and the elastic energy attains its minimum.

This stability analysis is confirmed by evaluating the sign definiteness of the Hessian of the elastic energy. For in-plane remodelling case here examined, convexity is equivalent to the positiveness of the term $H_{33}=\Hc[\Wb_3,\Wb_3]$, with $\Wb_3=\eb_1\otimes\eb_2-\eb_2\otimes\eb_1$. By using \eqref{hes} and by considering that the energy depends on $I_4$ only, as in \eqref{phi4phi5}, one obtains
\begin{equation}
H_{33}=(\lambda_1-\lambda_2)\big[2\,\phi_4\sin^2(\theta)-2\cos^2(\theta)\,\big(\phi_4+(\lambda_2-\lambda_1)\phi_{44}\sin^2(\theta)\big)\big]\,,
\end{equation} 
where $\phi_4$ and $\phi_{44}$ are evaluated from Eq.~\eqref{phi4phi5}.

In correspondence of the three possible stationary states of Eq.~\eqref{1dremod}, $\theta_\ast=0^\circ$, $\theta_\ast=90^\circ$, and $\theta_{\ast}$ corresponding to $\phi_4^\ast=0$, previous equation gives the following stability chart of the system:\\
\begin{center}
\begin{tabular}{l cccc}
\toprule
&\multicolumn{2}{c}{$\lambda_1>1$}&\multicolumn{2}{c}{$\lambda_1<1$}\\[0.1cm]
$\theta_\ast=0^\circ$ & $\lambda_1>\lambda_2$ & $\lambda_1<\lambda_2$ & $\lambda_1>\lambda_2$ &$\lambda_1<\lambda_2$\\[0.1cm]
&unstable & stable & stable & unstable\\[0.1cm]
\hline
&&&&\\[-0.3cm]
&\multicolumn{2}{c}{$\lambda_2>1$}&\multicolumn{2}{c}{$\lambda_2<1$}\\[0.1cm]
$\theta_\ast=90^\circ$ & $\lambda_2>\lambda_1$ & $\lambda_2<\lambda_1$ & $\lambda_2>\lambda_1$ &$\lambda_2<\lambda_1$\\[0.1cm]
& unstable & stable & stable & unstable\\[0.1cm]
\hline
&&&&\\[-0.3cm]
$\phi_4^\ast=0$& \multicolumn{4}{c}{stable for all $\lambda_1,\lambda_2$}\\
\bottomrule
\end{tabular}
\end{center}
that confirms the stability analysis based on the analysis of the derivatives of the function $\alpha(\theta)$ in Fig.~\ref{fig:alpha}.

\section{Traction-driven evolution of plane fibres}
In this section, we study the non-stationary solutions of the in-plane remodelling problem when the body is subjected to external loads. In particular, we consider a three--dimensional cube-like body $\Bc$, whose edges in the reference configuration are aligned along the directions of the orthonormal basis $\lbrace\eb_1, \eb_2, \eb_3\rbrace$, and whose faces are subjected to constant normal external loads $\pm\hat{s}_1$, $\pm\hat{s}_2$, $\pm\hat{s}_3$ measured as force per unit area in the reference configuration. 
The fibres are assumed to lie in the plane of unit normal $\eb_3$ and, with the aim to study an in-plane remodelling process, we further set $\hat s_3=0$ for all the loading cases examined. Correspondingly, the vector $\arem$ can be parametrised by the angle $\theta$, whose initial uniform value is defined by the scalar field $\theta_0=\theta_0(X)$. 

The strain energy function $\phi$ is assumed in the form of a neo-Hookean energy function, augmented by an anisotropic part only dependent on $I_4$:
\begin{equation}\label{enin}
\phi=\hat\phi(I_1,I_4)=\frac{1}{2}\mu\big(I_1-3 + \gamma(I_4-1)^2\big)\quad\textrm{and}\quad \det\Fb=1\,.
\end{equation}
As usual, enforcing the incompressibility constraint \eqref{enin}$_2$ makes mandatory to take into account also the reactive component of the stress and assume a representation of the reference stress $\Sb$ in the form: $\Sb = \hat\Sb(\Fb) - p\Fb^\star$, with the pressure $p$ as an additional state variable of the problem. 

In order to numerically solve the equilibrium problem \eqref{due}, the model was implemented in the commercial finite element software Comsol Multiphysics, which allowed us to solve the 1-parameter family of incompressible elastic problems, driven by the evolution of the orientation, in terms of the state variables $(\ub,p)$, together with the evolution problem of the fibres in terms of the  state variable $\theta$. To do that, the model was firstly posed in integral form as: find  $\ub$, $\theta$ and $p$ such that,
for all test fields $\tilde\ub$, $\tilde\theta$, $\tilde p$, the following condition holds true
\begin{equation}\label{bala_weak1}
0 =\displaystyle{\int_{\Bc} [\,-(\hat\Sb(\Fb)-p\,\Fb^\star)\cdot \nabla\tilde\ub\,]}+ \displaystyle{\int_{\Bc} [\,  (\dot\theta -\alpha(\theta)) \cdot\tilde\theta\,]}+\displaystyle{\int_{\Bc} [\,(\det\Fb  - 1) \cdot \tilde p}\,]+\displaystyle{\int_{\partial\Bc} [\,\hat{\bm{s}}\cdot\tilde{\ub}\,]}\,,
\end{equation}
where $\hat{\bm{s}}$ is the applied traction field on the boundary of the domain, $-p\Fb^\star$ is the reactive part of the reference stress $\Sb$, and $\alpha(\theta) = M(\theta)/m$ with $M(\theta)$ the Eshelby torque per unit of referential volume. In the in-plane case considered here, this latter quantity is given by the scalar field 
\begin{equation}
M(\theta)=\frac{1}{2}\mu\,\gamma\,(I_4-1)\bigl((C_{11}-C_{22})\sin(2\theta) -C_{12}\cos(2\theta)\bigr)\,.
\label{EshelbyTorque}
\end{equation}
Accordingly, the ratio $t_c=m/(\mu\gamma)$ defines the characteristic time of the evolution process, that depends on the shear modulus $\mu$, the stiffening parameter $\gamma$, and the mobility $m$.  
%
Within the finite element software, the cube-like body was discretized by tetrahedral elements with approximately $1300$ DOF. The numerical implementation of the full remodelling problem was carried out in two steps. Firstly, the solution $(\bar\ub(X),\bar p(X))$ of the purely elastic problem corresponding to the fiber-reinforced cube with a uniform orientation $\theta_0$ of the fibre was determined by means of a direct solver (MUMPS). Then, the full evolution problem was studied by using as initial values the solution of the previous step, i.e., $\theta(X,0)=\theta_0$, $\ub(X,0)=\bar\ub(X)$, and  $p(X,0)=\bar p(X)$, by means of a linear multistep method (BDF, order (5,1)). 

We studied three loading cases:
\begin{eqnarray}
\textrm{case I)}&&\hskip 1cm
\theta_0=10^\circ\quad\textrm{and}\quad
(\hat s_1,\hat s_2,\hat s_3)=\big(\frac{\mu}{2},-\frac{\mu}{4},0\big)\,,\\
\textrm{case II)}&&\hskip 1cm
\theta_0=40^\circ\quad\textrm{and}\quad
(\hat s_1,\hat s_2,\hat s_3)=\big(\frac{\mu}{2},\frac{3}{4}\mu,0\big)\,,\\
\textrm{case III)}&&\hskip 1cm
\theta_0=60^\circ\quad\textrm{and}\quad
(\hat s_1,\hat s_2,\hat s_3)=\big(\frac{\mu}{2},\frac{\mu}{4},0\big)\,,
\end{eqnarray}
that differ for the initial fibre orientation $\theta_0$ and for the ratio between the two normal tractions tractions $\pm\hat s_1$ and $\pm\hat s_2$. The values of the tractions were chosen to recover the three stationary regimes observed in the strain-controlled analysis of previous section; namely, one corresponding to a stationary orientation $0<\theta_\ast<90^\circ$ (case I), the other to $\theta_\ast=0^\circ$ (case II) and the third to $\theta_\ast=90^\circ$ (case III). It is noticed that in all the loading cases and at each time instant, the orientation $\theta$ as well as the strain and stress states are homogeneous. In all the three cases, the values of the constitutive parameters were $\mu=1$~KPa, $\gamma=1$ and $m=10$~Pa$\cdot$s, which correspond to a characteristic time of the remodelling process $t_c=10^{-2}$~s.
\begin{figure}[h]
\begin{center}
\begin{tiny}
 \def\svgwidth{\textwidth}
 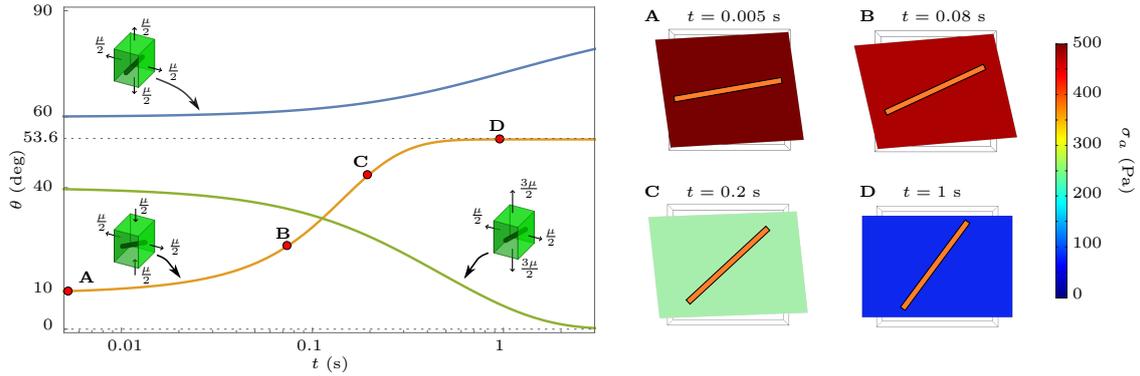
\end{tiny}
\end{center}
\caption{\label{fig:ev1} Time evolution of the fibre angle $\theta$ for the three loading cases: case I (orange curve), case II (green curve), and case III (blue curve) (semi-logarithmic plot). The green insets close to each curve represent the initial configuration of the cube and the corresponding tractions, whereas the contour plots show the stress along the fibre at selected time instants for the loading case I. For the blue curve, the simulation reached the value of 87.9$^\circ$ at $t=30$~s, but for the sake of clarity only the first 3~s are shown. 
}
\end{figure}

The results of the analysis in terms of the time evolution of the fibre angle are shown in Fig.~\ref{fig:ev1} in a semi-logarithmic plot (left panel). The colours of the curves are the same already used in Fig.~\ref{fig:alpha}: orange, green, and blue curves correspond to cases I, II, III, respectively. The green insets close to each curve represent the reference fibre-reinforced cube-like body and the corresponding tractions that drive the evolution of the fibre orientation from the initial value $\theta_0$ to the stationary value $\theta_\ast$. 
The contour plots in the same figure (right panel) show the evolution of the stress along the fibre, i.e., $\sigma_a=\bm\sigma\arem\cdot\arem$, at four time instants for the loading case I ($\theta_0=10^\circ$); the corresponding orientation of a representative fibre is also visible in the contour plots. As inferred from the analysis in the previous section, for such a loading case, the fibres will align in the direction for which $\phi_4=0$ (see Eq.~\eqref{implicitR}), thus $I_4 =1$ and the fibres do not experience any stretch; in the present case, this corresponds to an angle of about 53.6$^\circ$. Correspondingly, the stress $\sigma_a$ monotonically decreases during the remodelling process passing from about $500$~Pa, at the beginning of the evolution, to almost zero at the stationary solution; this latter value is not exactly zero due to the contribution of the isotropic part of the stress to $\sigma_a$. The analysis of strains at the stationary solutions further shows that in case II, for which $\theta_\ast=0^\circ$, the principal stretches are both compressive, with the one perpendicular to the fibre being lower than the other. On the contrary, in case III, for which $\theta_\ast=90^\circ$, both the stretches are tensile, and the one in the orthogonal direction is larger than the other. In this latter case, the simulation reached the value of 87.9$^\circ$ at $t=30$~s and tended asymptotically to $90^\circ$, but for the sake of clarity only the first 3~s are shown in Fig.~\ref{fig:ev1}.

\begin{figure}[h]
\begin{center}
\begin{tiny}
 \def\svgwidth{\textwidth}
 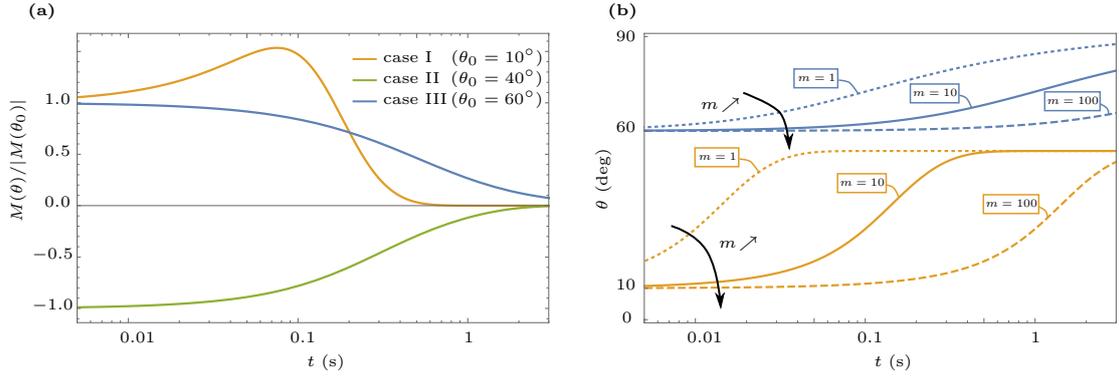
\end{tiny}
\end{center}
\caption{\label{fig:ev2} (a) Evolution of the Eshelby torque $M(t)=M(\theta(t))$ in the locading case I (orange curve), II (green curve), and III (blue curve). (b) Time evolution of $\theta$ for different values of the mobility parameter $m$ from $1$~Pa$\cdot$s to $100$~Pa$\cdot$s for loading cases I and III.}
\end{figure}

Figure~\ref{fig:ev2}a shows the evolution of the Eshelby torque $M$ in \eqref{EshelbyTorque} normalised to the its initial value $\vert M(\theta_0)\vert$; the Eshelby torque evolves as a consequence of the evolution of the fibre orientation $\theta$. Case I (orange curve in the figure) displays a non-monotonic behaviour with the torque initially increasing and then decreasing until vanishing at the stationary solution. This actually corresponds to the non-monotonic behaviour shown in Fig.~\ref{fig:alpha} for $\alpha(\theta)$ (we recall that $M(\theta)=m \,\alpha(\theta)$), although in that figure the stationary value of the angle was different due to the slightly different values of the principal stretches in the two analyses. A similar analysis shows that the Eshelby torque corresponding to case II (green curve) and case III (blue curve) evolves towards zero monotonically. In particular, the negative value of $M$ in case II (green curve) is due to the fact that the torque acts clockwise to make the fibre orientation evolving from the initial angle $\theta_0=40^\circ$ towards the stationary value $\theta_\ast=0^\circ$.

Finally, the effects of the mobility $m$  on the evolution process are shown in Fig.~\ref{fig:ev2}b for $m=\lbrace 1,10,100\rbrace$~Pa$\cdot$s; for the sake of clarity only two loading cases (I and III) are displayed. It is seen that an increase in the mobility parameter corresponds to an increase of the characteristic time of the evolution process that causes the fibres to reach their stationary orientation at higher times.

\section{Conclusions and future directions}
In this paper, a new model for transverse isotropic materials in which the reference orientation of fibres can evolve with time has been introduced within the context of finite elasticity with remodeling.  
The classical balance laws of elasticity are complemented with an evolution equation that controls the remodelling process, satisfying an extended energy imbalance principle. The inner source of the evolution equation is the Eshelbian torque, whereas the external source may incorporate stimuli driving the evolution of the fibres.  
In the passive case, when no external stimuli act on the material, fibres reorient themselves to make the inner torque vanishing. In such a configuration, the stress and strain tensors are coaxial and the elastic energy has an extremal point. The corresponding stationary orientation of the fibre may be aligned with a principal strain direction, but other orientations are indeed possible.
 These additional configurations have been completely characterised by showing that they are the solutions of an implicit equation of the invariants $I_4$ and $I_5$: if this equation admits a solution, the corresponding remodelled orientation of the fibres represents a stable configuration, i.e., a minimum of the elastic energy; in this case, the principal strain directions, that are still stationary solutions of the evolution equation, are indeed maxima of the elastic energy, thus unstable configurations.
On the contrary, if the implicit equation does not have a solution, the fibres will align along a principal strain direction, with the stable configuration being associated with the direction of minimum principal strain.
The model has been implemented in Comsol Multiphysics to simulate the fibre reorientation in a mixed boundary value problem and can be used to study the reorientation of active materials or biological tissues.

The results of our analysis show that the strain-driven or stress-driven approaches followed in the literature, in which the fibres are aligned along the maximum principal strain or stress directions, are only possible in the presence of external stimuli and by supplementing energy to the system. Yet, these approaches can be easily integrated in our model by introducing a suitable forcing term in the evolution equation. Likewise, the thermodynamic framework used to derive the remodelling equation can be extended to study the residual stress effects which arise from the fibre reorientation driven by external stimuli. Viscoelastic effects can also be accounted for by properly modifying the constitutive equation of the stress tensor.

\section*{Competing interests}
The authors declare no competing interests.

\section*{Acknowledgements}
This publication is based on the work supported by Sapienza Universit\`a di Roma under the project ''Mechanics of soft fibered active materials'' (N. RG11715C7CE2C1C4).
%


\begin{thebibliography}{10}
\expandafter\ifx\csname url\endcsname\relax
  \def\url#1{\texttt{#1}}\fi
\expandafter\ifx\csname urlprefix\endcsname\relax\def\urlprefix{URL }\fi
\expandafter\ifx\csname href\endcsname\relax
  \def\href#1#2{#2} \def\path#1{#1}\fi

\bibitem{Taber:1996}
L.~A. Taber, D.~W. Eggers,
  \href{http://www.sciencedirect.com/science/article/pii/S0022519396901071}{Theoretical
  study of stress-modulated growth in the aorta}, Journal of Theoretical
  Biology 180~(4) (1996) 343 -- 357.
\newblock \href {http://dx.doi.org/https://doi.org/10.1006/jtbi.1996.0107}
  {\path{doi:https://doi.org/10.1006/jtbi.1996.0107}}.
\newline\urlprefix\url{http://www.sciencedirect.com/science/article/pii/S0022519396901071}

\bibitem{Cowin:2004}
S.~C. Cowin,
  \href{https://doi.org/10.1146/annurev.bioeng.6.040803.140250}{Tissue growth
  and remodeling}, Annual Review of Biomedical Engineering 6~(1) (2004)
  77--107, pMID: 15255763.
\newblock \href
  {http://arxiv.org/abs/https://doi.org/10.1146/annurev.bioeng.6.040803.140250}
  {\path{arXiv:https://doi.org/10.1146/annurev.bioeng.6.040803.140250}}, \href
  {http://dx.doi.org/10.1146/annurev.bioeng.6.040803.140250}
  {\path{doi:10.1146/annurev.bioeng.6.040803.140250}}.
\newline\urlprefix\url{https://doi.org/10.1146/annurev.bioeng.6.040803.140250}

\bibitem{Garikipati:2006}
K.~Garikipati, J.~Olberding, H.~Narayanan, E.~Arruda, K.~Grosh, S.~Calve,
  \href{http://www.sciencedirect.com/science/article/pii/S002250960600007X}{Biological
  remodelling: Stationary energy, configurational change, internal variables
  and dissipation}, Journal of the Mechanics and Physics of Solids 54~(7)
  (2006) 1493 -- 1515.
\newblock \href {http://dx.doi.org/https://doi.org/10.1016/j.jmps.2005.11.011}
  {\path{doi:https://doi.org/10.1016/j.jmps.2005.11.011}}.
\newline\urlprefix\url{http://www.sciencedirect.com/science/article/pii/S002250960600007X}

\bibitem{Alford:2007}
P.~W. Alford, J.~D. Humphrey, L.~A. Taber,
  \href{https://doi.org/10.1007/s10237-007-0101-2}{Growth and remodeling in a
  thick-walled artery model: effects of spatial variations in wall
  constituents}, Biomechanics and Modeling in Mechanobiology 7~(4) (2007) 245.
\newblock \href {http://dx.doi.org/10.1007/s10237-007-0101-2}
  {\path{doi:10.1007/s10237-007-0101-2}}.
\newline\urlprefix\url{https://doi.org/10.1007/s10237-007-0101-2}

\bibitem{Criscione:2007}
J.~C. Criscione, \href{https://doi.org/10.1007/s10237-007-0103-0}{Kinematics
  framework optimized for deformation, growth, and remodeling in vascular
  organs}, Biomechanics and Modeling in Mechanobiology 7~(4) (2007) 285.
\newblock \href {http://dx.doi.org/10.1007/s10237-007-0103-0}
  {\path{doi:10.1007/s10237-007-0103-0}}.
\newline\urlprefix\url{https://doi.org/10.1007/s10237-007-0103-0}

\bibitem{Nagel:2012}
T.~Nagel, D.~J. Kelly, Remodelling of collagen fibre transition stretch and
  angular distribution in soft biological tissues and cell-seeded hydrogels.,
  Biomechanics and modeling in mechanobiology 11 3-4 (2012) 325--39.

\bibitem{NematNasser:2002}
S.~Nemat-Nasser, \href{https://doi.org/10.1063/1.1495888}{Micromechanics of
  actuation of ionic polymer-metal composites}, Journal of Applied Physics
  92~(5) (2002) 2899--2915.
\newblock \href {http://arxiv.org/abs/https://doi.org/10.1063/1.1495888}
  {\path{arXiv:https://doi.org/10.1063/1.1495888}}, \href
  {http://dx.doi.org/10.1063/1.1495888} {\path{doi:10.1063/1.1495888}}.
\newline\urlprefix\url{https://doi.org/10.1063/1.1495888}

\bibitem{Galante:2013}
S.~Galante, A.~Lucantonio, P.~Nardinocchi,
  \href{http://www.sciencedirect.com/science/article/pii/S0020746213000073}{The
  multiplicative decomposition of the deformation gradient in the multiphysics
  modeling of ionic polymers}, International Journal of Non-Linear Mechanics 51
  (2013) 112 -- 120.
\newblock \href
  {http://dx.doi.org/https://doi.org/10.1016/j.ijnonlinmec.2013.01.005}
  {\path{doi:https://doi.org/10.1016/j.ijnonlinmec.2013.01.005}}.
\newline\urlprefix\url{http://www.sciencedirect.com/science/article/pii/S0020746213000073}

\bibitem{Dicarlo:2002}
A.~DiCarlo, S.~Quiligotti,
  \href{http://www.sciencedirect.com/science/article/pii/S0093641302002975}{Growth
  and balance}, Mechanics Research Communications 29~(6) (2002) 449 -- 456.
\newblock \href
  {http://dx.doi.org/https://doi.org/10.1016/S0093-6413(02)00297-5}
  {\path{doi:https://doi.org/10.1016/S0093-6413(02)00297-5}}.
\newline\urlprefix\url{http://www.sciencedirect.com/science/article/pii/S0093641302002975}

\bibitem{Tiero:2016}
A.~Tiero, G.~Tomassetti, \href{https://doi.org/10.1177/1081286514546178}{On
  morphoelastic rods}, Mathematics and Mechanics of Solids 21~(8) (2016)
  941--965.
\newblock \href {http://arxiv.org/abs/https://doi.org/10.1177/1081286514546178}
  {\path{arXiv:https://doi.org/10.1177/1081286514546178}}, \href
  {http://dx.doi.org/10.1177/1081286514546178}
  {\path{doi:10.1177/1081286514546178}}.
\newline\urlprefix\url{https://doi.org/10.1177/1081286514546178}

\bibitem{Rodriguez:1994}
E.~K. Rodriguez, A.~Hoger, A.~D. McCulloch,
  \href{http://www.sciencedirect.com/science/article/pii/0021929094900213}{Stress-dependent
  finite growth in soft elastic tissues}, Journal of Biomechanics 27~(4) (1994)
  455 -- 467.
\newblock \href
  {http://dx.doi.org/https://doi.org/10.1016/0021-9290(94)90021-3}
  {\path{doi:https://doi.org/10.1016/0021-9290(94)90021-3}}.
\newline\urlprefix\url{http://www.sciencedirect.com/science/article/pii/0021929094900213}

\bibitem{Nardinocchi:2012}
P.~Nardinocchi, L.~Teresi, V.~Varano,
  \href{http://www.sciencedirect.com/science/article/pii/S0022509612000828}{Strain
  induced shape formation in fibred cylindrical tubes}, Journal of the
  Mechanics and Physics of Solids 60~(8) (2012) 1420 -- 1431.
\newblock \href {http://dx.doi.org/https://doi.org/10.1016/j.jmps.2012.04.010}
  {\path{doi:https://doi.org/10.1016/j.jmps.2012.04.010}}.
\newline\urlprefix\url{http://www.sciencedirect.com/science/article/pii/S0022509612000828}

\bibitem{Efrati:2013}
E.~Efrati, E.~Sharon, R.~Kupferman,
  \href{http://dx.doi.org/10.1039/C3SM50660F}{The metric description of
  elasticity in residually stressed soft materials}, Soft Matter 9 (2013)
  8187--8197.
\newblock \href {http://dx.doi.org/10.1039/C3SM50660F}
  {\path{doi:10.1039/C3SM50660F}}.
\newline\urlprefix\url{http://dx.doi.org/10.1039/C3SM50660F}

\bibitem{Pezzulla:2016}
M.~Pezzulla, G.~P. Smith, P.~Nardinocchi, D.~P. Holmes,
  \href{http://dx.doi.org/10.1039/C6SM00246C}{Geometry and mechanics of thin
  growing bilayers}, Soft Matter 12 (2016) 4435--4442.
\newblock \href {http://dx.doi.org/10.1039/C6SM00246C}
  {\path{doi:10.1039/C6SM00246C}}.
\newline\urlprefix\url{http://dx.doi.org/10.1039/C6SM00246C}

\bibitem{Goriely:2017}
A.~Goriely, The Mathematics and Mechanics of Biological Growth,
  Interdisciplinary Applied Mathematics 45, Springer, 2017.

\bibitem{Aharoni:2017}
H.~Aharoni, J.~M. Kolinski, M.~Moshe, I.~Meirzada, E.~Sharon,
  \href{https://link.aps.org/doi/10.1103/PhysRevLett.117.124101}{Internal
  stresses lead to net forces and torques on extended elastic bodies}, Phys.
  Rev. Lett. 117 (2016) 124101.
\newblock \href {http://dx.doi.org/10.1103/PhysRevLett.117.124101}
  {\path{doi:10.1103/PhysRevLett.117.124101}}.
\newline\urlprefix\url{https://link.aps.org/doi/10.1103/PhysRevLett.117.124101}

\bibitem{Neville:1993}
A.~C. Neville, Biology of fibrous composites : development beyond the cell
  membrane, Cambridge University Press, 1993.

\bibitem{Driessen:2003}
N.~Driessen, Remodelling of continuously distributed collagen fibres in soft
  connective tissues, Journal of Biomechanics 36~(8) (2003) 1151 -- 1158.
\newblock \href
  {http://dx.doi.org/https://doi.org/10.1016/S0021-9290(03)00082-4}
  {\path{doi:https://doi.org/10.1016/S0021-9290(03)00082-4}}.

\bibitem{Cardamone:2009}
L.~Cardamone, A.~Valent{\'i}n, J.~F. Eberth, J.~D. Humphrey,
  \href{https://doi.org/10.1007/s10237-008-0146-x}{Origin of axial prestretch
  and residual stress in arteries}, Biomechanics and Modeling in Mechanobiology
  8~(6) (2009) 431.
\newblock \href {http://dx.doi.org/10.1007/s10237-008-0146-x}
  {\path{doi:10.1007/s10237-008-0146-x}}.
\newline\urlprefix\url{https://doi.org/10.1007/s10237-008-0146-x}

\bibitem{Ericksen:1991}
J.~Ericksen, Introduction to the Thermodynamics of Solids, Chapman \& Hall,
  1991.

\bibitem{deGennes:1995}
P.~G. de~Gennes, J.~Prost, Biology of fibrous composites : development beyond
  the cell membrane, Clarendon Press, 1995.

\bibitem{Sebastian:2018}
N.~Sebastián, N.~Osterman, D.~Lisjak, M.~Čopič, A.~Mertelj,
  \href{http://dx.doi.org/10.1039/C8SM01377B}{Director reorientation dynamics
  of ferromagnetic nematic liquid crystals}, Soft Matter 14 (2018) 7180--7189.
\newblock \href {http://dx.doi.org/10.1039/C8SM01377B}
  {\path{doi:10.1039/C8SM01377B}}.
\newline\urlprefix\url{http://dx.doi.org/10.1039/C8SM01377B}

\bibitem{DeSimone:2007}
A.~DeSimone, A.~DiCarlo, L.~Teresi,
  \href{https://doi.org/10.1140/epje/i2007-10240-2}{Critical voltages and
  blocking stresses in nematic gels}, The European Physical Journal E 24~(3)
  (2007) 303.
\newblock \href {http://dx.doi.org/10.1140/epje/i2007-10240-2}
  {\path{doi:10.1140/epje/i2007-10240-2}}.
\newline\urlprefix\url{https://doi.org/10.1140/epje/i2007-10240-2}

\bibitem{Fukunaga:2008}
A.~Fukunaga, K.~Urayama, T.~Takigawa, A.~DeSimone, L.~Teresi,
  \href{https://pubs.acs.org/doi/abs/10.1021/ma801639j}{Dynamics of
  electro-opto-mechanical effects in swollen nematic elastomers},
  Macromolecules 41~(23) (2008) 9389 -- 9396.
\newblock \href {http://dx.doi.org/10.1021/ma801639j}
  {\path{doi:10.1021/ma801639j}}.
\newline\urlprefix\url{https://pubs.acs.org/doi/abs/10.1021/ma801639j}

\bibitem{Sawa:2010}
Y.~Sawa, K.~Urayama, T.~Takigawa, A.~DeSimone, L.~Teresi,
  \href{https://pubs.acs.org/doi/abs/10.1021/ma1003979}{Thermally driven giant
  bending of liquid crystal elastomer films with hybrid alignment},
  Macromolecules 43~(9) (2010) 4362 -- 4369.
\newblock \href {http://dx.doi.org/10.1021/ma1003979}
  {\path{doi:10.1021/ma1003979}}.
\newline\urlprefix\url{https://pubs.acs.org/doi/abs/10.1021/ma1003979}

\bibitem{Stanier:2016}
D.~C. Stanier, J.~Ciambella, S.~S. Rahatekar,
  \href{http://www.sciencedirect.com/science/article/pii/S1359835X16303281}{Fabrication
  and characterisation of short fibre reinforced elastomer composites for
  bending and twisting magnetic actuation}, Composites Part A: Applied Science
  and Manufacturing 91 (2016) 168 -- 176.
\newblock \href
  {http://dx.doi.org/https://doi.org/10.1016/j.compositesa.2016.10.001}
  {\path{doi:https://doi.org/10.1016/j.compositesa.2016.10.001}}.
\newline\urlprefix\url{http://www.sciencedirect.com/science/article/pii/S1359835X16303281}

\bibitem{Ciambella:2017}
J.~Ciambella, D.~C. Stanier, S.~S. Rahatekar,
  \href{http://www.sciencedirect.com/science/article/pii/S1359836816317206}{Magnetic
  alignment of short carbon fibres in curing composites}, Composites Part B:
  Engineering 109~(Supplement C) (2017) 129 -- 137.
\newblock \href
  {http://dx.doi.org/https://doi.org/10.1016/j.compositesb.2016.10.038}
  {\path{doi:https://doi.org/10.1016/j.compositesb.2016.10.038}}.
\newline\urlprefix\url{http://www.sciencedirect.com/science/article/pii/S1359836816317206}

\bibitem{Ciambella:2017_PRSA}
J.~Ciambella, A.~Favata, G.~Tomassetti,
  \href{http://rspa.royalsocietypublishing.org/content/474/2209/20170703}{A
  nonlinear theory for fibre-reinforced magneto-elastic rods}, Proceedings of
  the Royal Society of London A: Mathematical, Physical and Engineering
  Sciences 474~(2209).
\newblock \href
  {http://arxiv.org/abs/http://rspa.royalsocietypublishing.org/content/474/2209/20170703.full.pdf}
  {\path{arXiv:http://rspa.royalsocietypublishing.org/content/474/2209/20170703.full.pdf}},
  \href {http://dx.doi.org/10.1098/rspa.2017.0703}
  {\path{doi:10.1098/rspa.2017.0703}}.
\newline\urlprefix\url{http://rspa.royalsocietypublishing.org/content/474/2209/20170703}

\bibitem{Dicarlo:2006}
A.~DiCarlo, S.~Naili, S.~Quiligotti,
  \href{http://www.sciencedirect.com/science/article/pii/S1631072106001112}{Sur
  le remodelage des tissus osseux anisotropes}, Comptes Rendus Mécanique
  334~(11) (2006) 651 -- 661.
\newblock \href {http://dx.doi.org/https://doi.org/10.1016/j.crme.2006.06.009}
  {\path{doi:https://doi.org/10.1016/j.crme.2006.06.009}}.
\newline\urlprefix\url{http://www.sciencedirect.com/science/article/pii/S1631072106001112}

\bibitem{Himpel:2008}
G.~Himpel, A.~Menzel, E.~Kuhl, P.~Steinmann,
  \href{https://onlinelibrary.wiley.com/doi/abs/10.1002/nme.2124}{Time‐dependent
  fibre reorientation of transversely isotropic continua—finite element
  formulation and consistent linearization}, International Journal for
  Numerical Methods in Engineering 73~(10) (2008) 1413--1433.
\newblock \href
  {http://arxiv.org/abs/https://onlinelibrary.wiley.com/doi/pdf/10.1002/nme.2124}
  {\path{arXiv:https://onlinelibrary.wiley.com/doi/pdf/10.1002/nme.2124}},
  \href {http://dx.doi.org/10.1002/nme.2124} {\path{doi:10.1002/nme.2124}}.
\newline\urlprefix\url{https://onlinelibrary.wiley.com/doi/abs/10.1002/nme.2124}

\bibitem{Melnik:2013}
A.~V. Melnik, A.~Goriely,
  \href{https://doi.org/10.1177/1081286513485773}{Dynamic fiber reorientation
  in a fiber-reinforced hyperelastic material}, Mathematics and Mechanics of
  Solids 18~(6) (2013) 634--648.
\newblock \href {http://arxiv.org/abs/https://doi.org/10.1177/1081286513485773}
  {\path{arXiv:https://doi.org/10.1177/1081286513485773}}, \href
  {http://dx.doi.org/10.1177/1081286513485773}
  {\path{doi:10.1177/1081286513485773}}.
\newline\urlprefix\url{https://doi.org/10.1177/1081286513485773}

\bibitem{Menzel:2005}
A.~Menzel, \href{https://doi.org/10.1007/s10237-004-0047-6}{Modelling of
  anisotropic growth in biological tissues}, Biomechanics and Modeling in
  Mechanobiology 3~(3) (2005) 147--171.
\newblock \href {http://dx.doi.org/10.1007/s10237-004-0047-6}
  {\path{doi:10.1007/s10237-004-0047-6}}.
\newline\urlprefix\url{https://doi.org/10.1007/s10237-004-0047-6}

\bibitem{Kuhl:2005}
E.~Kuhl, K.~Garikipati, E.~M. Arruda, K.~Grosh, {Remodeling of biological
  tissue: Mechanically induced reorientation of a transversely isotropic chain
  network}, Journal of the Mechanics and Physics of Solids 53~(7) (2005)
  1552--1573.
\newblock \href {http://arxiv.org/abs/0411037} {\path{arXiv:0411037}}, \href
  {http://dx.doi.org/10.1016/j.jmps.2005.03.002}
  {\path{doi:10.1016/j.jmps.2005.03.002}}.

\bibitem{Hariton:2007}
I.~Hariton, G.~DeBotton, T.~C. Gasser, G.~A. Holzapfel, {Stress-driven collagen
  fiber remodeling in arterial walls}, Biomechanics and Modeling in
  Mechanobiology 6~(3) (2007) 163--175.
\newblock \href {http://dx.doi.org/10.1007/s10237-006-0049-7}
  {\path{doi:10.1007/s10237-006-0049-7}}.

\bibitem{Driessen:2008}
N.~J. Driessen, M.~A. Cox, C.~V. Bouten, F.~P. Baaijens, {Remodelling of the
  angular collagen fiber distribution in cardiovascular tissues}, Biomechanics
  and Modeling in Mechanobiology 7~(2) (2008) 93--103.
\newblock \href {http://dx.doi.org/10.1007/s10237-007-0078-x}
  {\path{doi:10.1007/s10237-007-0078-x}}.

\bibitem{Grillo:2018}
A.~Grillo, M.~Carfagna, S.~Federico,
  \href{https://doi.org/10.1007/s10665-017-9940-8}{An allen--cahn approach to
  the remodelling of fibre-reinforced anisotropic materials}, Journal of
  Engineering Mathematics 109~(1) (2018) 139--172.
\newblock \href {http://dx.doi.org/10.1007/s10665-017-9940-8}
  {\path{doi:10.1007/s10665-017-9940-8}}.
\newline\urlprefix\url{https://doi.org/10.1007/s10665-017-9940-8}

\bibitem{Merodio:2005}
J.~Merodio, R.~Ogden,
  \href{http://www.sciencedirect.com/science/article/pii/S002074620400054X}{Mechanical
  response of fiber-reinforced incompressible non-linearly elastic solids},
  International Journal of Non-Linear Mechanics 40~(2) (2005) 213 -- 227,
  special Issue in Honour of C.O. Horgan.
\newblock \href
  {http://dx.doi.org/https://doi.org/10.1016/j.ijnonlinmec.2004.05.003}
  {\path{doi:https://doi.org/10.1016/j.ijnonlinmec.2004.05.003}}.
\newline\urlprefix\url{http://www.sciencedirect.com/science/article/pii/S002074620400054X}

\bibitem{Coleman:1974}
B.~D. Coleman, W.~Noll, \href{https://doi.org/10.1007/978-3-642-65817-4_9}{The
  Thermodynamics of Elastic Materials with Heat Conduction and Viscosity},
  Springer Berlin Heidelberg, Berlin, Heidelberg, 1974, pp. 145--156.
\newblock \href {http://dx.doi.org/10.1007/978-3-642-65817-4_9}
  {\path{doi:10.1007/978-3-642-65817-4_9}}.
\newline\urlprefix\url{https://doi.org/10.1007/978-3-642-65817-4_9}

\bibitem{Vianello:1996a}
M.~Vianello, \href{https://doi.org/10.1007/BF00042131}{Optimization of the
  stored energy and coaxiality of strain and stress in finite elasticity},
  Journal of Elasticity 44~(3) (1996) 193--202.
\newblock \href {http://dx.doi.org/10.1007/BF00042131}
  {\path{doi:10.1007/BF00042131}}.
\newline\urlprefix\url{https://doi.org/10.1007/BF00042131}

\end{thebibliography}

%
\end{document}